\documentclass[12pt]{article}
\usepackage{apacite}
\usepackage[utf8]{inputenc}
\usepackage{amsmath,amssymb}
\usepackage{setspace}
\usepackage[margin=1in]{geometry}
\usepackage{natbib}
\usepackage{url}
\usepackage{hyperref}
\usepackage{enumitem}
\usepackage{amsmath,amssymb,amsfonts}
\usepackage{algorithmic}
\usepackage{textcomp}
\usepackage{xcolor}
\usepackage{lipsum}
\usepackage{listings}
\newtheorem{definition}{Definition}
\newtheorem{insight}{Insight}
\usepackage{graphicx,color,url} % Required for inserting images
\usepackage{ltablex}
\usepackage{booktabs} % Allows the use of \toprule, \midrule and \bottomrule in tables
\usepackage{multirow}
\usepackage[table]{xcolor}
\usepackage{kotex}

\title{Quantifying the Engagement Effectiveness of Cyber Cognitive Attacks: \\A Behavioral Metric for Disinformation Campaigns}

\author{Bonnie Rushing and Shouhuai Xu, Ph.D.\\
University of Colorado Colorado Springs \\
\texttt{brushing@uccs.edu}}
\date{\today}

\begin{document}

\title{Quantifying the Engagement Effectiveness of Cyber Cognitive Attacks:\\A Behavioral Metric for Disinformation Campaigns}

\maketitle

\begin{abstract}
As disinformation-driven cognitive attacks become increasingly sophisticated, the ability to quantify their impact is essential for advancing cybersecurity defense strategies. This paper presents a novel framework for measuring the engagement effectiveness of cognitive attacks by introducing a weighted interaction metric that accounts for both the type and volume of user engagement relative to the number of attacker-generated transmissions. Applying this model to real-world disinformation campaigns across social media platforms, we demonstrate how the metric captures not just reach but the behavioral depth of user engagement. Our findings provide new insights into the behavioral dynamics of cognitive warfare and offer actionable tools for researchers and practitioners seeking to assess and counter the spread of malicious influence online.
\end{abstract}


\section{Highlights}
\begin{itemize}
    \item Introduces a weighted metric to quantify effectiveness of cyber cognitive attacks.
    \item Develops a grading scale to classify disinformation content by impact.
    \item Applies the metric to real-world cases across multiple platforms.
    \item Extends existing frameworks (e.g., DISARM) with measurable engagement assessment.
    \item Enables automation and future use in real-time influence detection.
\end{itemize}
\textit{Disclaimer: “The views expressed are those of the author and do not reflect the official policy or position of the US Air Force Academy, US Air Force, Department of Defense, or the US Government.”}

\section{Introduction}

\textit{Cyber cognitive attacks}, including disinformation, aim to use computer-enabled technologies to advance psychological battles into victims' subconsciousness, adversely affecting human security and privacy online. Strategic cognitive attacks threaten personal security worldwide through victims' engagement with malicious disinformation efforts. Cognitive attackers aim to achieve objectives such as reshaping perceptions or harming democratic processes and international security \citep{NATOscience,blacksea}.

The full scope of cognitive operations and their resulting social implications is unclear and complex because attacks include advanced technologies and impact neuroscience and psychology to maximize \textit{effectiveness}. Cognitive warfare operations rapidly increase in quantity, range, precision, and \textit{effectiveness}, increasing their potential for strategic success \citep{USNI}. This threat is growing and affects populations globally--it is vital to understand and measure the impacts of cognitive warfare.

This research aims to develop a novel approach to measuring the effectiveness of cognitive warfare. We analyze current methods and build on existing state-of-the-art research. The goal is to provide means to calculate quantifiable metrics for cyber cognitive attack engagements. This metric will enhance understanding of cyber cognitive attacks and introduce a method for researchers and practitioners to quantify attacks.

Discovering related work and existing data on cognitive attackers and their effects is essential for measuring security impacts. For example, we must understand what constitutes an ``\textit{effective} cognitive attack'' and what response, if any, is required from victims to be affected by attackers' efforts.

Cognitive warfare operations, and information operations in general, need quantifiable cybersecurity metrics for measurement and comparison. This research aims to provide a model to measure the \textit{engagement effectiveness} of cyber cognitive attacks and share findings to advance information warfare research and frameworks.

\subsection{Contributions}
This paper makes four original contributions, including 
(i) an original definition of effective cyber cognitive attacks, highlighting how attackers achieve their objectives by exposing victims to content like disinformation online; 
(ii) the creation of a novel methodology that introduces a metric to quantify cyber cognitive attack engagement effectiveness, which measures effectiveness as the weighted sum of user interactions divided by the number of attacker transmissions, providing a graded output of attack success; 
(iii) an empirical evaluation of the metric application to real-world case studies, highlighting metric functionality and insights with three alleged cyber cognitive attacks and publicly available content on Instagram, YouTube, and Facebook;  
(iv) and original recommendations for future research in this field including proposals for validating this paper's metric, analyzing its application temporally, and suggestions to fill other gaps within the Disinformation Analysis and Response Measures (DISARM) framework \citep{DISARM}.

\subsection{Problem Statement Details}

This section defines this project's problem statement and associated cyber cognitive attack terminology.

\subsubsection{Problem statement}
This paper aims to solve the following problems and answer these questions. (i) What is cyber cognitive attack engagement? (ii) Why is cognitive attack effectiveness significant? (iii) Are there attempts to measure cognitive attack effectiveness? (iv) What is the state-of-the-art? 
Our literature review shows no such disinformation measurements have been published or standardized. Perhaps the research community has not realized the importance of standardizing cognitive attack metrics and language and has not prioritized measurement efforts. The community will benefit from a practical solution to quantify this growing technological threat. 

\textit{Problem Statement}: Despite the increasing prevalence of cognitive attacks using disinformation to manipulate perceptions, there is currently no standardized method to quantify their engagement effectiveness. This lack of quantifiable metrics impedes cybersecurity efforts to identify and mitigate these attacks in real-time. How can we develop a reliable and scalable metric to measure the engagement effectiveness of cognitive attacks across diverse digital platforms? 

\subsubsection{Cyber Cognitive Attack Background}
Cognitive attack operations, enhanced through the weaponization of users' personal information and technology, transmit content to target victims' subliminal cognition, including their underlying emotions, knowledge, willpower, and beliefs. The objective involves affecting targets' cognition, attitude, and behavior through subtle influence \citep{RAND}. Tactics for cyber cognitive attacks include developing disinformation campaign designs, goals, content, and accounts, driving online harms, delivering content, persisting in the information environment, and maximizing exposure \citep{DISARM}. A PLA Daily article argues that subliminal messaging acts as a carrier to transmit visual, auditory, and other information that people are unaware of but can cause subliminal reactions, enabling intervention and control of human cognition, emotion, and will. Subliminal information can bypass human consciousness and act directly on the unconscious as covert psychological operations \citep{RAND}. 
Further, the PLA expressed interest in progressing research on hidden psychological warfare operations in foreign countries, unconscious control, and the ability to influence victims' willingness to accept indoctrination, public opinion warfare, and propaganda. Additionally, with increased network transmission speeds and algorithm technology, cyber cognitive attackers can quickly insert subliminal information into many files and upload them to popular platforms, making cognitive attacks hard to guard against \citep{RAND}.
Based on this data, we deduce that in the case of cognitive attacks' subliminal effects, no conscious response is required from victims to be affected by attackers' efforts. 

\begin{definition}
    [Cyber cognitive attack] Online operations that target human minds' subconsciousness, aiming to manipulate perceptions and beliefs, which may be weaponized and enhanced through technology and deceptive information, typically to affect individuals' or broader populations' decision-making and actions to gain advantages. Cognitive attacks need not be overly sophisticated to achieve desired effects. Instead, cognitive attack effects may be achieved subliminally.
\end{definition}

Subconscious effects are subliminal impacts transmitted through visual, auditory, and other information that people are unaware of. They lead to cognitive reactions, intervention, and control of human cognition, emotion, and will. Subliminal effects can bypass human consciousness and act directly on the unconscious as covert psychological operations \citep{RAND}. Thus, measuring cognitive attack effectiveness is complex and challenging. 

\subsubsection{Cyber Cognitive Attack Attributes}
\begin{figure}
    \centering
    \includegraphics[width=.5\linewidth]{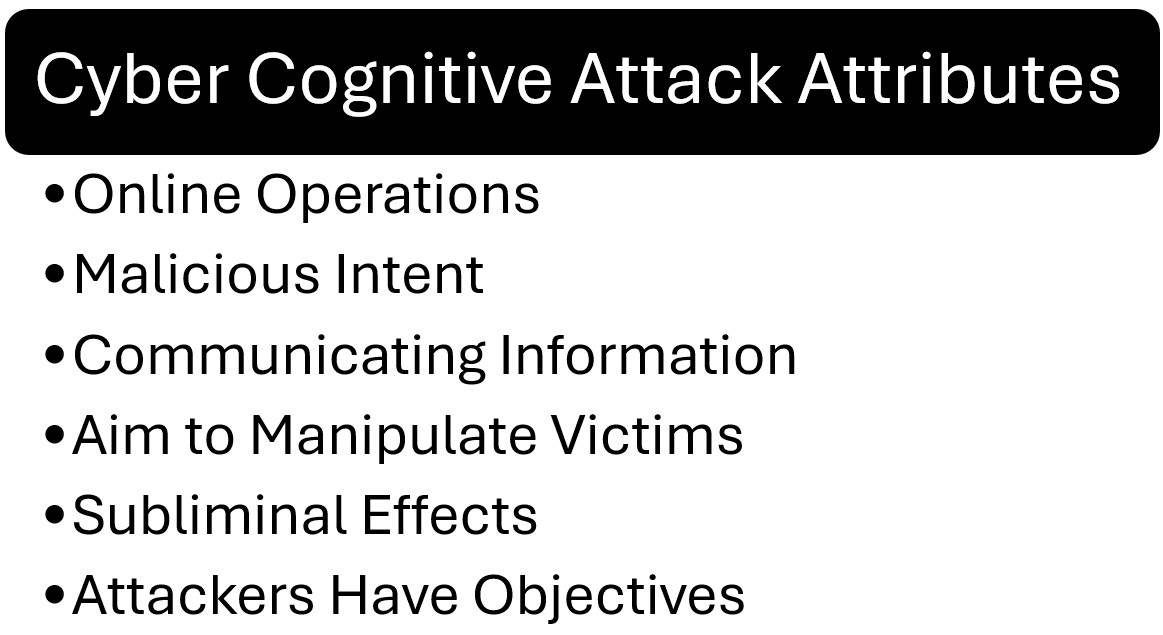}
    \caption{Cyber Cognitive Attack Attributes}
    \label{fig:attributes}
\end{figure}

Investigators shall use the following descriptions when selecting case study data for our methodology. We present the following distinct attributes of cyber cognitive attacks in connection with the terminology defined in this paper:
\begin{itemize}
    \item Online operations - \textit{cyber} cognitive attacks refer to operations on the Internet or other networked technologies \citep{RAND}; 
    \item Malicious intent - cyber cognitive attacks intentionally drive online harm through transmitted content like disinformation \citep{RAND}; 
    \item Communicating information - cyber cognitive attacks include varied communication forms through audio, visual, and other mediums via online platforms \citep{USNI}; 
    \item Aim to manipulate victims' subconsciousness -  attacks expose victims to malicious content, aiming to influence perceptions, beliefs, and opinions to meet attackers' objectives \citep{RAND}\citep{blacksea}\citep{USNI}. 
\end{itemize}
Investigators may consider these attributes essential when selecting transmissions for analysis using our methodology. We consider these qualities to be inherently part of cyber cognitive attacks, depicted in Figure \ref{fig:attributes}.

\subsubsection{Engagement Effectiveness}
This paper aims to measure cognitive attack \textit{engagement effectiveness}, meaning that a metric may be applied to grade cognitive attacks based on observable \textit{engagements (interactions)} with online cognitive attack content. Thus, this paper defines effective cognitive attacks as follows.

\begin{definition}
[Effective cyber cognitive attack] A cyber cognitive attack that leads to the achievement of the attacker's objectives by reaching and exposing the target audience to selected messaging, even for a split second. 
\end{definition}

\begin{definition}
[Cyber cognitive attack Engagement] Engagement refers to quantified user interactions rather than measuring how attackers effectively meet their objectives. The engagement with the attacker's content indicates their level of success, evidenced by visible interaction metrics.

Let $E$ denote the engagement effectiveness of a cognitive attacker's efforts. Let $t$ represent the total number of cognitive warfare-related transmissions, such as disinformation posts, Deepfake images or videos, and malicious messages.

Let $i_j$ represent the number of interactions of type $j$, and $w_j$ the associated weight assigned to each interaction type based on its relative cognitive or behavioral significance. Specifically, we assign:
\begin{itemize}
    \item View ($w_{\text{view}}$ = 0.1)
    \item Like ($w_{\text{like}}$ = 0.3)
    \item Comment ($w_{\text{comment}}$ = 0.7)
    \item Share ($w_{\text{share}}$ = 1.0)
\end{itemize}

We define the total weighted interaction score $I$ as:
\[
I = \sum_{j=1}^{n} w_j \cdot i_j
\]

We then define engagement effectiveness $E$ as:
\[
E = \frac{I}{t}
\]

This weighted metric reflects both the volume and the qualitative depth of user interactions, addressing the limitations of prior formulations that treated all interaction types as equal. The weighting scheme is informed by prior work in social media engagement modeling \citep{STIEGLITZ2018156} and user attention studies \citep{krishnan2013videoads}.

\end{definition}

The details of this mathematical formula are explained in Section \ref{section:method}. The effectiveness of cyber cognitive attacks refers to reaching and engaging online audiences, making engagement interactions a valuable metric for understanding cyber cognitive attacks. In \textit{cognitive} warfare, effectiveness may be achieved with the user's split-second view or repetitive exposure of content to embed mental biases. For example, subliminal exposure to a stimulus increases associated positive feelings about it \citep{exposure}. 

\subsubsection{Actors}

\textit{Cyber Cognitive Attackers}—or \textit{cognitive attackers} for short. These actors represent the aggressors in the cyber cognitive attack threat model. Cyber cognitive attackers transmit malicious communications, such as disinformation, designed to meet their objectives through cognitive influence.

\textit{Users} or \textit{Victims}—Internet users who may encounter cyber cognitive attacks. In the cyber cognitive attack threat model, users represent vulnerable targets and may be influenced by content such as disinformation online.

\subsubsection{Metrics}
The metrics are defined as follows.

\begin{definition}
[grade]  A metric representing the cyber cognitive attacker's engagement effectiveness, measuring the attack's success as evidenced by user interactions.
\end{definition}

\begin{definition}
[interactions] The users' actions when engaging with online content (e.g., likes, comments, saves, views, reads, plays, and shares on posts, videos, and other content) are used in the metric calculation.
\end{definition}

\begin{definition}
[transmissions] the quantity of cognitive warfare-related content transmitted, e.g., disinformation posts, Deepfake images or videos, and messages. Transmissions are used in the metric calculation.
\end{definition}

\subsection{Project Objectives}

This project primarily seeks to develop a method to measure the effectiveness of social media engagement in cyber cognitive attacks. Thus, this paper's analysis aims to answer this primary research question:

\textbf{Research Question:} \textbf{How should we quantify the engagement effectiveness of cognitive attacks?} 

To answer this question, we delve into the following inquiries on how ``cognitive attack engagement effectiveness'' should be measured.
What do we need to measure (metrics), and how should it be defined? For example, should we consider (i) the quantity of transmitted content (e.g., cognitive attacker's social media posts and messages), (ii) quantities of visible or recordable reactions and interactions with cyber cognitive attack content (e.g., likes, shares, and comments), or (iii) quantities of all interactions with attackers' online content, including user views?
    
How can we quantify (measure) cognitive attack capabilities' effectiveness? For example, should we consider (i) mathematical calculations of attackers' content interactions, or (ii) cognitive attacker's online account reach and following?

How can we empirically or experimentally validate or invalidate these metrics? For example, should we consider (i) victims' actions in the physical world? (e.g., talking about cognitive attack details or making decisions based on cognitive attack effects) or (ii) a victim's psychological factors, or attackers' psychological tactics and techniques employed \citep{sophistication}?

Based on this analysis, we will introduce a model to calculate the engagement effectiveness of cyber cognitive attacks and evaluate case studies to exemplify the model's functionality. Finally, we present recommendations for future research to build on this paper and the current state-of-the-art.

\subsection{Paper Outline}

This paper analyzes these existing data and related work in Section \ref{section:related} to discover applicable state-of-the-art statuses and defining relevant terminology. Further, this project's methodology in  Section \ref{section:method} introduces the novel metric to quantify cyber cognitive warfare effectiveness. Next, we apply the model to real-world case studies, highlighting functionality and findings in \ref{section:findings}. We highlight research and application recommendations in Section \ref{section:discussion}. Finally, we conclude this project in  Section \ref{section:conclusion}.

\subsection{Related Work}
\label{section:related}

Cognitive warfare effectiveness is not a new research topic. However, most studies appear qualitative in the current state-of-the-art. We believe that this paper introduces the first cyber cognitive attack engagement metrics to the community.

This project seeks to build knowledge within the DISARM Framework \citep{DISARM}, which categorizes disinformation incidents, seeking to fight disinformation internationally by sharing data, analysis, and coordinating action. This framework is essential because it was developed by drawing on global cybersecurity best practices. It is used to help interdisciplinary communicators gain a clear shared understanding of disinformation incidents and to identify defensive actions. The framework was constructed based on historical and hypothetical tactics and techniques employed by manipulators (attackers) and responses employed by defenders \citep{DISARM}. However, its state-of-the-art frameworks do not contain datasets, detection methods, counters, incidents, or examples for the following items in their Tactic stage: TA12, \textit{Assess Effectiveness}, including:
\begin{itemize}
    \item \textit{T0134.002 ``Social media engagement'' - Summary: Monitor and evaluate social media engagement in misinformation incidents.}
\end{itemize}
The lack of data for \textit{T0134.002, ``Social media engagement,''} motivates this research to develop a method to measure the effectiveness of social media engagement in cyber cognitive attacks. 

Few quantitative or measurement studies are available on cyber cognitive attack effects, but researchers have published relevant metrics. These metrics included Internet users' self-reported disinformation encounters \citep{stats}. Self-reported metrics may not be sufficient because most people are unaware they have been exposed to cognitive attacks. Therefore, most victims may not be able to provide accurate feedback on cognitive warfare's effects on them \citep{swords}. 
Nevertheless, researchers started introducing quantitative measurements to the community. A sophistication framework was defined and evaluated based on the study, "Quantifying Psychological Sophistication of Malicious Emails \citep{sophistication}," which investigated the psychological sophistication of short malicious emails. It has two pillars: Psychological Techniques (PTechs) and Psychological Tactics (PTacs). The study proposes metrics to assess the sophistication of malicious emails through PTechs and PTacs.

%Psychology researchers indicate that effects may be achieved in \textit{cognitive} warfare with as little as a victim's split-second view or repetitive exposure to content to embed mental biases. For example, subliminal exposure to a stimulus increases associated positive feelings about the content. Sophistication may not be required \citep{exposure}. This subliminal messaging is explained further in a RAND research paper \citep{RAND}, which highlights how subliminal messaging uses information as a carrier to transmit visual, auditory, and other information that victims are not aware of but can cause subliminal reactions, aiming to control human cognition, emotion, and will.

Research on disinformation engagement data is vastly qualitative, including a paper by Matina Rapti, George Tsakalidis, Sophia Petridou, and Kostas Vergidis \citep{info13070306}. In their paper, their researchers claim that the content recipients (victims) are the individuals who use the Internet as a portal to receive the news but also engage with content, repost articles, react, comment, ``like,'' and share without being the content producers (attackers) themselves. Content recipients can propagate disinformation by reading, relating to, and sharing a story they encountered online without previously cross-checking the accuracy of the content \citep{info13070306}. Rapti's project studies real-world disinformation incidents and categorizes them qualitatively into deceptive patterns. This disinformation causes international security risks.

For this paper, we assign the following weighting logic to engagements: \textit{View = 0.1, Like = 0.3, Comment = 0.7, Share = 1.0}. These are \textit{illustrative approximations} informed by both academic literature and platform behavior heuristics.

Stieglitz et al. \citep{STIEGLITZ2018156} propose a tiered model of interaction significance, categorizing engagement types by user effort and potential influence:
\begin{itemize}
    \item Low-effort: views, likes
    \item Moderate-effort: comments
    \item High-effort, high-diffusion: shares
\end{itemize}
While their framework does not define explicit numeric weights, it supports the hierarchical structure, which underpins our chosen ordering: \begin{verbatim}share > comment > like > view
\end{verbatim}

Krishnan and Sitaraman \citep{krishnan2013videoads} reinforce this approach through a large-scale measurement study on video ad effectiveness. Their findings suggest that deeper interactions—such as sustained attention or click-through behavior—are better indicators of user engagement and intent, aligning with our increased weighting of comments and shares.

To understand effective cognitive attacks and the resulting security implications, we analyze the closest related work by RAND, which states that PLA (People's Liberation Army) researchers use subliminal messaging to circumvent victims' consciousness and resistance to receiving undesired information \citep{RAND}. The PLA intelligent warfare strategy involves this subliminal messaging during psychological warfare content creation and transmission and impacts victims' internal processing \citep{RAND}. 
  
\section{Methodology}
\label{section:method}
\begin{figure}
    \centering
    \includegraphics[width=0.3\linewidth]{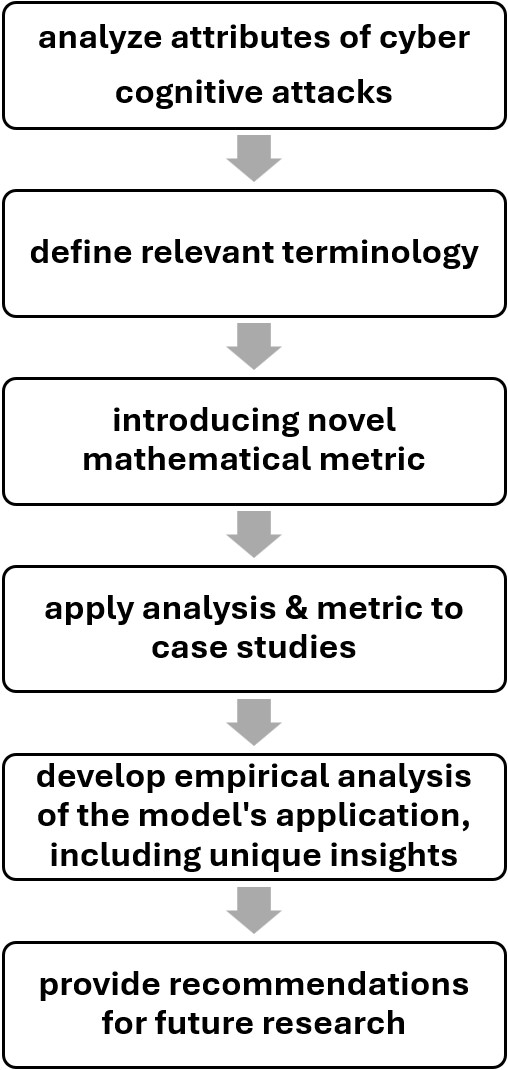}
    \caption{Methodology Flow Chart}
    \label{fig:flowchart}
\end{figure}

To quantify cognitive attack engagement effectiveness, we propose a methodology that will guide our case studies and may be applied to other research in future studies. One unique feature of this methodology is that cyber cognitive attack engagements are elusive to measure, meaning that a competent methodology should adequately consider the state-of-the-art literature. Our methodology has the following steps: (i) analyzing attributes of cyber cognitive attacks, (ii) defining relevant terminology, (iii) introducing the novel mathematical metric, (iv) applying the analysis and metric to case studies, (v) developing an empirical analysis of the model's application, including unique insights, and (vi) providing recommendations for future research. Figure \ref{fig:flowchart} depicts this methodology's flow chart.

\subsection{Metric}

To address our core research question, \textit{``How should we quantify the engagement effectiveness of cognitive attacks?''}, this paper introduces a novel metric based on weighted user engagement. The goal is to quantify not just the volume but the qualitative significance of user interactions with cognitive attack content.

Let $E$ denote the engagement effectiveness of a cognitive attacker's efforts. Let $t$ represent the number of cognitive warfare-related transmissions (e.g., disinformation posts, Deepfake images, or deceptive videos).

Let $i_j$ represent the number of interactions of type $j$, and $w_j$ be the assigned weight for each interaction type based on cognitive significance. Specifically, we use:

\begin{itemize}
    \item View ($w_{\text{view}}$ = 0.1)
    \item Like ($w_{\text{like}}$ = 0.3)
    \item Comment ($w_{\text{comment}}$ = 0.7)
    \item Share ($w_{\text{share}}$ = 1.0)
\end{itemize}

We define the total weighted interaction score $I$ as:
\[
I = \sum_{j=1}^{n} w_j \cdot i_j
\]

We then define engagement effectiveness $E$ as:
\[
E = \frac{I}{t}
\]

This approach distinguishes between passive and active engagement signals, addressing prior limitations that treated all interactions equally. The weights are informed by behavioral theory and prior work in social media interaction modeling \citep{STIEGLITZ2018156, krishnan2013videoads}. The resulting metric is scalable and adaptable to different platforms and can help assess the depth and potential impact of cognitive attack efforts.

The $E$ score is then interpreted using our ordinal grading scale in Table~\ref{tab:grades}, which classifies content from \textit{failure} to \textit{viral} based on interaction intensity per transmission.

\subsubsection{Model Measurement Justification}
This model prioritizes the effectiveness of cyber cognitive attack engagement rather than other measurements, such as effectively meeting the attacker's objectives, their quantity of posts, or users' specific actions. For example, even if troll farms publish thousands of disinformation posts daily, the large quantity of transmissions \textit{($t$)} alone does not make these efforts \textit{effective}. Further, these posts are not guaranteed to be delivered to users to produce desired cognitive attack effects. By contrast, if a cognitive attacker transmits one post, it may be effective by reaching users whose interactions are used as evidence for its \textit{engagement effectiveness} in this case. 

The specific \textit{engagement} verbiage refers to successfully quantified user engagement interactions rather than measuring how attackers effectively meet their objectives. The engagement with the attacker's content indicates their level of success, evidenced by interaction metrics.
In cognitive warfare, a victim may only need to ``glance'' or listen passively (e.g., reaching a quantifiable number of video or post views or play counts as interactions \textit{($i$)}) to consume split-second messaging and be affected by the attack subliminally \citep{exposure}. Therefore, we propose that no specific reactions or factors from victims are required to validate a cognitive attack's engagement effectiveness. These performance parameters indicate how successfully the disinformation reaches users without explicitly measuring how well it meets the attacker's mission objectives.

In the realm of cognitive attacks, engagement metrics such as views, likes, shares, and comments are clear indicators of the attack’s reach and the audience’s interaction with the content. The underlying logic is that higher engagement correlates with a greater chance that the attack has succeeded in influencing perceptions, opinions, or behavior.
Research on cognitive warfare has shown that even short exposures to disinformation can affect beliefs or reinforce biases \citep{exposure}. Thus, by measuring user interactions, the model captures the subliminal impact of repeated or subtle cognitive stimuli (e.g., split-second exposure to disinformation), which can be challenging to quantify otherwise. Using publicly available engagement data is a pragmatic choice because it provides an observable and measurable aspect of cognitive attacks. It enables a standardized approach to analyzing attacks across various platforms, even when direct measurement of cognitive effects (e.g., psychological changes in users) is not feasible.

 The metric offers a simple, scalable, and repeatable measurement method. This is crucial for cybersecurity professionals who require a tool that can be easily applied across multiple platforms without demanding complex computational resources. This metric is adaptable to different online environments (e.g., social media, blogs, and video platforms), making it highly versatile and easy to implement with real-time data. Our model prioritizes quantity and virality over quality, which is crucial for understanding the dissemination and exposure of disinformation in cognitive attacks. 

The following metric definitions serve as a starting point and may require further refinement, particularly as technologies evolve and cyber interactions continue to develop. To our knowledge, no previously established metrics for cyber attack engagement effectiveness of this type exist, and our grading scale is based on the following logic. 
This methodology considers available and visible interactions from the originator, estimating that cognitive attackers may contribute two interactions (e.g., liking and commenting) on their content. The DISARM framework categorizes this technique as \textit{Tactic stage: TA09, Delivering content by replying or commenting via owned media (assets that the operator controls)} \citep{DISARM}. Thus, we define attack failure as content without a single view or interaction from a user. 
\begin{itemize}
    \item \textit{Grade F, a \textit{failure}, considers two interactions from the attacker or related account(s). \textit{(E is 2 or less)}.}
    \item Based on this logic, we increase interaction minimums for each higher grade.
    \item \textit{Grade E}, \textit{poor}, considers two interactions from the attacker or related account(s) plus one or two additional interactions. (\textit{E is 3 to 4}) In essence, this grade assumes at least one user's interaction.
    \item \textit{Grade D}, \textit{below average}, considers two interactions from the attacker or related account(s) plus two to seven additional interactions. \textit{(E is 4 to 9) In essence, this grade assumes less than ten user interactions.}
    \item \textit{Grade C}, \textit{average}, considers two interactions from the attacker or related account(s) plus at least eight additional interactions. (\textit{E is 10 to 99}) In essence, this grade assumes less than 100 user interactions.
    \item \textit{Grade B}, \textit{above average}, considers two interactions from the attacker or related account(s) plus at least 98 additional interactions. (\textit{E is 100 to 999})  In essence, this grade assumes less than a thousand user interactions.
    \item \textit{Grade A}, \textit{excellent}, considers two interactions from the attacker or related account(s) plus at least 998 additional interactions. (\textit{E is 1,000 to 9,999}) In essence, this grade assumes less than ten thousand user interactions.
    \item \textit{Grade A}+, \textit{viral}, considers two interactions from the attacker or related account(s) plus at least 9,998 additional interactions. (\textit{E is 10,000+}) In essence, this grade involves more than ten thousand user interactions.
At the \textit{A+} effectiveness grade, we categorize the associated media content as \textit{viral disinformation}. Based on the literature review conducted for this paper, there is no standardized metric to categorize online viral content. Thus, we propose that the viral content measurement should be a minimum of 10,000 interactions. For every multiple of 10,000 \textit{($E$)}, we describe the content as multiples, e.g., viral x2, viral x3.
\end{itemize}

This cyber cognitive attack effectiveness grading scale interpretation is depicted in Table \ref{tab:grades}\footnote{Grade thresholds are based on the illustrative weights defined in our engagement effectiveness metric. As these weights are not empirically derived but grounded in theoretical and behavioral frameworks, the thresholds may require recalibration if future work adopts alternative weighting schemes or integrates additional interaction types.}.

\subsubsection{Model Limitations}

While the proposed weighted interaction metric addresses several shortcomings of earlier formulations, it is not without limitations.

First, the assigned weights for different interaction types (e.g., views, likes, shares) are approximations based on behavioral theory and prior literature \citep{STIEGLITZ2018156, krishnan2013videoads}, rather than empirical fitting to cognitive influence outcomes. As such, the metric reflects a normative model of engagement rather than a validated predictive measure of psychological or behavioral change.

Second, the metric does not currently distinguish between unique users. A single actor engaging with content multiple times (e.g., bots or hyperactive accounts) could inflate the weighted interaction score. Future work should incorporate user-level deduplication or bot filtering methods to mitigate this effect.

Third, the model assumes each transmission has equal strategic intent and potential reach. In practice, some transmissions (e.g., a viral Deepfake or coordinated hashtag campaign) may carry significantly more influence than others. Integrating message-level virality or platform-specific amplification could improve fidelity.

Fourth, the current metric is static and does not yet account for the temporal evolution of engagement, such as repeated exposure or delayed diffusion. Future extensions could model time-series engagement patterns to reflect longitudinal cognitive influence.

Finally, certain interactions such as "plays" and "saves" are currently excluded due to their ambiguous cognitive meaning—i.e., they may not signal belief, endorsement, or diffusion. Including these requires a more nuanced analysis, potentially informed by platform affordances and user intent models.

\begin{table}
    \centering
    \begin{tabular}{|c|c|l|} \hline 
         Effectiveness ($E$)&  Grade Scale  &Description\\ \hline 
         10,000+&  A+ &Viral\\ \hline 
         1,000 - 9,999&  A &Excellent\\ \hline 
         100 - 999&  B &Above Average\\ \hline 
         10 - 99&  C &Average\\ \hline 
         4 - 9&  D &Below Average\\ \hline 
         3 - 4&  E &Poor\\ \hline 
         0 - 2&  F &Failure\\\hline
    \end{tabular}
    \caption{Grading Scale for Cyber Cognitive Attack Engagement Effectiveness (E)}
    \label{tab:grades}
\end{table}

\begin{figure*}
\centering
  \includegraphics[width=1\textwidth,height=7cm]{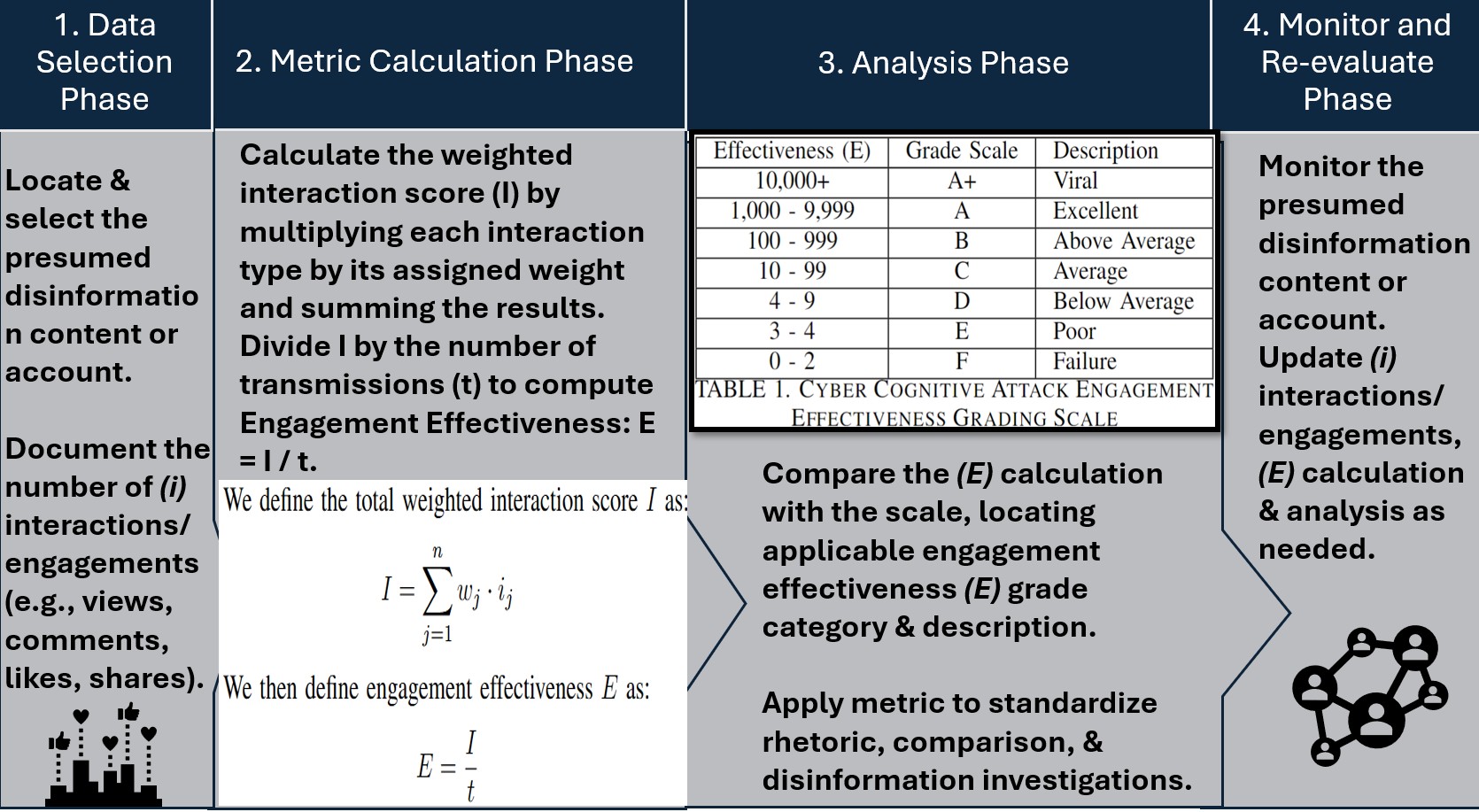}
  \caption{Scientific Process Overview for Cyber Cognitive Attack Engagement Effectiveness Calculation and Analysis}
  \label{fig:process}
\end{figure*}

\subsection{Overview of Scientific Process}

The metric's application should be calculated using the scientific process according to the following four phases:\\
\textit{1. Data Selection Phase} - Locate and select the presumed disinformation content or account online. Document the number of publicly visible \textit{($i$)} interactions with the selected disinformation (e.g., views, comments, likes, shares). Social media analytics platforms may assist in data collection.\\

\textit{2. Metric Calculation Phase} - Multiply each interaction type by its assigned weight, sum the weighted values to obtain the total interaction score $I$, and divide by the number of transmissions $t$ to compute engagement effectiveness $E = I / t$.

\textit{3. Analysis Phase -} Compare the \textit{($E$)} calculation with the ordinal scale in Table \ref{tab:grades}, recording applicable engagement effectiveness \textit{($E$)} grade category and description.
Apply the metric to standardize interdisciplinary rhetoric, data comparison, and disinformation investigations.\\

\textit{4. Monitor and Re-evaluate Phase} - Monitor the originally selected disinformation content or account at time intervals that suit mission requirements as needed. Update the associated \textit{($i$)} interactions and \textit{($E$)} calculation \& analysis as needed.\\

Figure \ref{fig:process} depicts this scientific process overview for calculating and analyzing cyber cognitive attack engagement effectiveness.

\section{Case Studies}
\label{section:findings}

This section applies the metric methodology to three case studies of alleged cyber cognitive attacks, highlighting its functionality. These case studies and related data collection and screenshots include publicly available content and visible interaction data from three websites: Instagram, YouTube, and Facebook. This section also includes proposed future research recommendations based on these findings and state-of-the-art developments.

\begin{figure}[htbp]
    \centering
    \includegraphics[width=.4\linewidth]{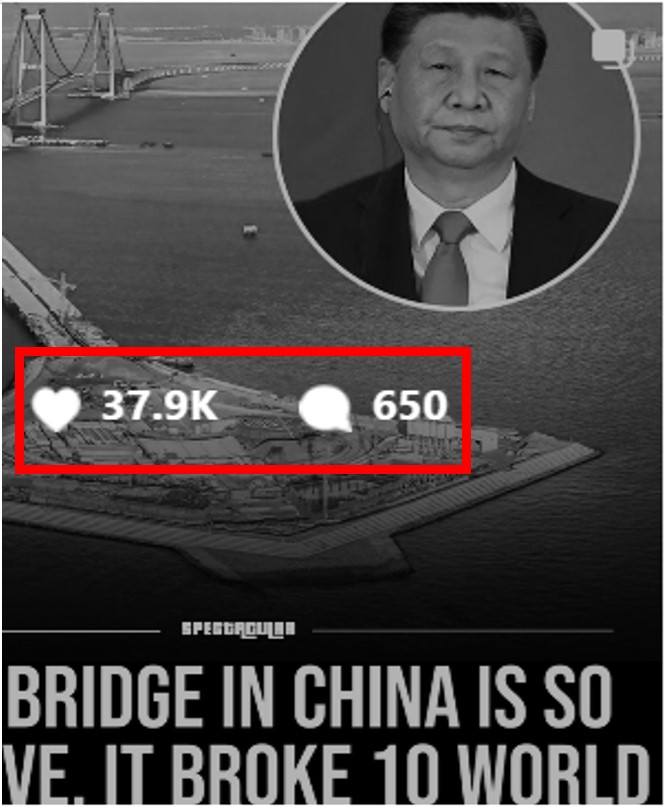}
   \caption{Case Study 1, Instagram Post Interactions}
    \label{fig:insta}
\end{figure}

\subsection{Case Study 1: Instagram}
In July 2024, a Pro-China Instagram post grew popular with engagement activity through an account called \textit{@spectacular}. The post content involved China’s Shenzhen-Zhongshan Link in Guangdong. The original content creator boldly claims this so-called 'landmark in engineering' sets ten world records. Notably, these positive statements and photos cannot be corroborated through reputable sources online as of September 2024.
This paper's methodology considers this single post as an example of one (quantity \textit{($t$)=1}) cyber cognitive attack. 

Based on publicly available information on September 18, 2024, we found 37,908 likes and 650 comments on this post, as shown in Figure \ref{fig:insta}.

The post received 37,908 likes and 650 comments, visible in Figure~\ref{fig:insta}. Using our weighted interaction metric, the effectiveness score ($E$) is calculated as:

\[
E = \frac{(0.3 \cdot 37,908) + (0.7 \cdot 650)}{1} = 11,827.4
\]

Based on the Cyber Cognitive Attack Engagement Effectiveness Grading Scale in Table~\ref{tab:grades}, this post is categorized as \textit{A+}, viral (x1).

\begin{figure}[htbp]
    \centering
    \includegraphics[width=.7\linewidth]{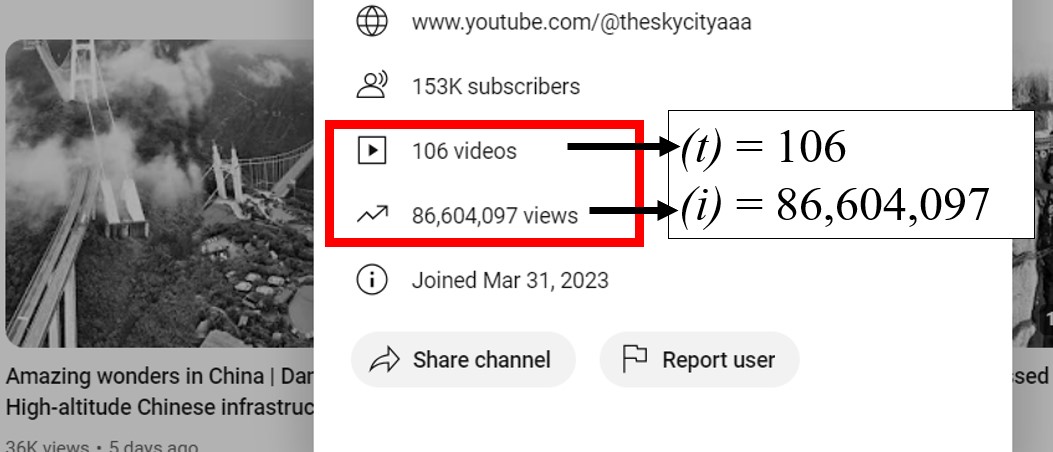}
    \caption{Case Study 2, YouTube Account Video Views as Interactions}
    \label{fig:videoexample}
\end{figure}

\subsection{Case Study 2: YouTube}
A screenshot captured on September 19, 2024 in Figure \ref{fig:videoexample} depicts a presumed pro-China YouTube account, \textit{@theskycityaaa}. This YouTube account contains 106 videos, uploaded within about 1.5 years. Most videos depict a bias toward positive aspects of life in China, including content titles that highlight ``Amazing Wonders in China,'' ``Amazing Chinese Infrastructure,'' ``China's Infrastructure Miracle,'' ``Americans are Stunned,'' and ``China's Cliff Wonders.''

We grade the whole account overview instead of an individual post for this example case. The content summary revealed 106 video uploads \textit{($t$)} and 86,604,097 views \textit{($i$)}. Further, investigation teams may also decide to include individual likes and comments for each of the 106 videos as needed. 

The account summary revealed 106 video uploads \textit{($t$)} and 86,604,097 views. Using the weighted interaction metric with views assigned a weight of 0.1:

\[
E = \frac{0.1 \cdot 86,604,097}{106} = \frac{8,660,409.7}{106} \approx 81,700.1
\]

This places the account in the \textit{A+} category, viral (x8), based on Table~\ref{tab:grades}. A more comprehensive analysis could further enhance accuracy by incorporating likes, comments, and shares for each video.

\subsection{Case Study 3: Facebook}
\begin{figure}[htbp]
    \centering
    \includegraphics[width=.7\linewidth]{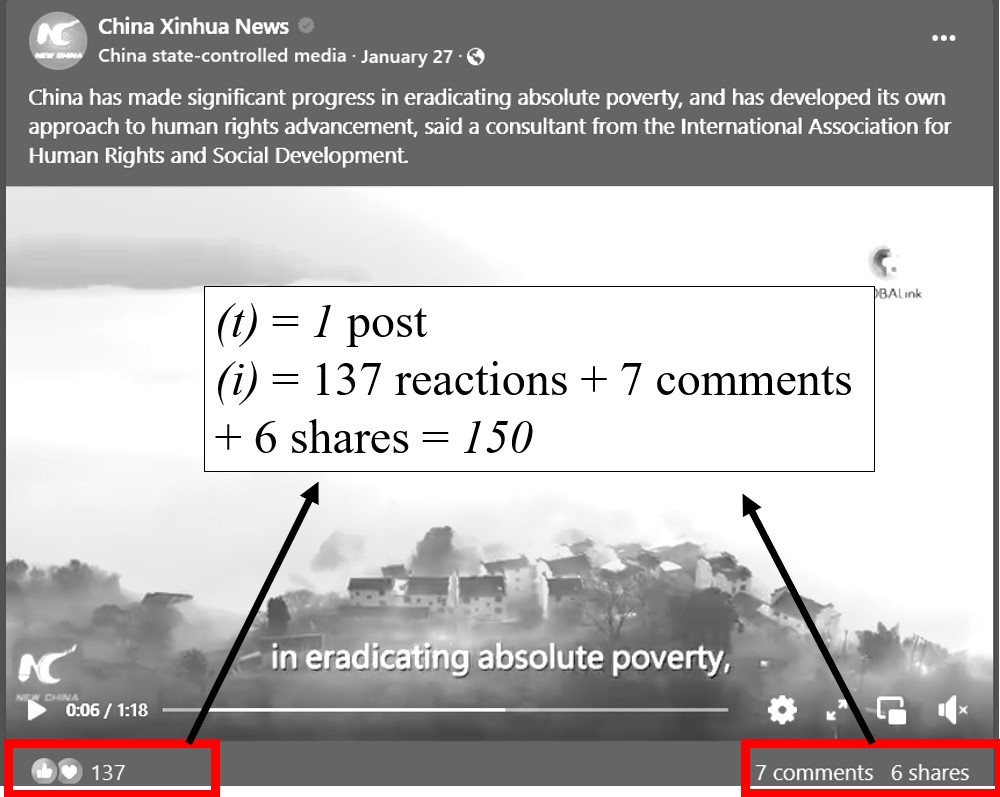}
    \caption{Case Study 3, Facebook Video Post with Diverse Interactions}
    \label{fig:chinanews}
\end{figure}

A screenshot captured on September 20, 2024 in Figure \ref{fig:chinanews} depicts content from a Chinese state-controlled media outlet, \textit{China Xinhua News}. This example analyzes their video post on Facebook, dated January 27, 2024. This single post \textit{(t = 1)} generated 137 reactions (likes), 7 comments, and 6 shares. Applying the weights:

\[
E = \frac{(0.3 \cdot 137) + (0.7 \cdot 7) + (1.0 \cdot 6)}{1} = 52.0
\]

According to Table~\ref{tab:grades}, this effectiveness score corresponds to a grade of \textit{C}, average.

\section{Discussion}
\label{section:discussion}

The engagement effectiveness metric ($E$) represents an effort to better align quantification with the cognitive impact of disinformation campaigns. By assigning weights to interaction types based on user effort and diffusion potential, the metric distinguishes between passive exposure (e.g., views) and active dissemination (e.g., shares), which better mirrors the intent and objectives of cognitive warfare actors.

This approach also addresses reviewer concerns that the original metric was overly simplistic and susceptible to false positives. Still, the metric should be seen as a foundational scaffold rather than a definitive measure. Future extensions could incorporate user uniqueness, recursive exposure (second-order diffusion), or even longitudinal behavioral traces to model susceptibility and influence over time. These enhancements would further align the metric with current understandings of cognitive vulnerability and belief formation in disinformation research.

This research has profound implications for global security, the fight against disinformation, and future projects as cognitive warfare evolves. 
Investigators may use this metric to categorize cyber cognitive attacks, organize offensive data, use standardized measurements for comparison and discussion, and further understand the reach of cognitive warfare. This standardizes language and measurements, bringing interdisciplinary researchers into the same conversation. This metric may be applied across any online platform, including \textit{Twitter (X), TikTok,} or \textit{Reddit}, or analyzing attacks in different contexts (e.g., political campaigns, military operations, or corporate disinformation). Use cases include government counter-disinformation campaigns, social media monitoring systems, or corporate security policies. These examples show the flexibility and applicability of our model across various settings. 
Specifically, this metric introduces standardized measurements to help evaluate cognitive attack incidents, building tools to address gaps in cybersecurity frameworks, such as the DISARM Framework:
T0134.002 ``Social media engagement'' - Summary: Monitor and evaluate social media engagement in misinformation incidents." Our metric contributes to Key Performace Indicator (KPI) identification and tracking so that the performance and effectiveness of cognitive attack campaign elements can be measured during and after their execution  \citep{DISARM}.\\

\begin{insight}
\textit{The metric may be applied to aid and benefit such inquiries as:}
\begin{itemize}
    \item What is the average engagement effectiveness ($E$) of disinformation posts from certain accounts?
    \item How is cyber cognitive attack engagement connected to victims' real-world behavior?
    \item What type of accounts yield the highest engagement effectiveness ($E$) and why?
    \item How do known state-controlled media platforms compare on the metric grading scale?
    \item How do the engagement effectiveness ($E$) trends evolve with cyber cognitive attack accounts over time?
    \item Which disinformation incidents and accounts should be categorized as a threat to society or human security or flagged for removal from online platforms?
    \item How do government-backed disinformation campaign engagement trends develop? (real-world validation and contribution to policy-making)
\end{itemize}
\end{insight}

These issues may be solved with accompanying metric data indicating how the engagement effectiveness of the alleged disinformation incidents has impacted victims. This metric provides quantifiable measurements to vastly qualitative problems across cross-disciplinary fields. For example. cognitive attacks and their measurement intersect with many disciplines, like neuroscience, psychology, and information theory. The metric may also be automated online, providing investigators and cybersecurity agencies with real-time measurements of disinformation content engagement trends, which may prove advantageous during critical events like elections or crises. Furthermore, this metric is not only for measurement but also for active defense against cognitive attacks. For instance, the model could identify when attacks reach critical engagement levels, triggering mitigation efforts like content removal or user warnings and adding cyber resilience strategies. 
Users can automate the calculation of disinformation engagement effectiveness, apply a grade to the content, and prompt investigators to flag it for review (e.g., if the grade is ``A" or higher). A script may be coded to read engagement data (from an Application Programming Interface (API) or file), calculate the engagement effectiveness using our methodology, apply a grade, and empower users to track or mitigate disinformation online quickly. The code displayed in the Appendix demonstrates an example of this metric automation implementation in Python, assuming the client already has the engagement data for demonstration simplicity. 

\subsection{Recommended Future Research}

The field has knowledge and practitioner gaps, a lack of shared understanding and metrics of cyber cognitive attacks, and unified lines of effort. Based on our discoveries from developing this project's metric, analysis, and application insights, we present requirements that can guide the design of future research and solutions to understand and defeat these attacks. Cyber cognitive attack studies need formal requirements to optimize efforts and develop defenses. We propose the following research directions based on related studies and escalating security threats.

\begin{enumerate}
   \item Test validity by applying the metric to real-world or experimental cognitive attacks (e.g., comparative neuroscience metrics).
  \item If cognitive attackers' tactics and techniques are more sophisticated \citep{sophistication}, are they equally more effective in the metric?
  \item How may the cyber cognitive attack threat model change over time? 
  \item How may increased exposure time to cyber cognitive attacks affect victims?
 \item How can we quantify the effectiveness of cognitive defenses?
   \item The field would benefit from controlled and ethical human subjects research to discover the influence of cognitive attacks.
   \item The DISARM framework requires further development in this research field, including the gaps in T0133.001 Behavior changes, T0134.002 Social media engagement, T0133.003 Awareness, and T0133.005 Action/attitude \citep{DISARM}.

\end{enumerate}

Such research pathways will benefit international human security by growing the collective defense and understanding of cyber cognitive attacks.

\section{Conclusion}
\label{section:conclusion}
This paper introduced a novel metric for quantifying the engagement effectiveness of cyber cognitive attacks. By calculating the ratio of weighted user interactions to the number of transmissions, our metric offers a standardized and scalable approach for measuring the impact of disinformation campaigns across various social media platforms. Through detailed case studies on Instagram, YouTube, and Facebook, we demonstrated the applicability and utility of this metric in understanding how cognitive attacks engage with and influence their target audiences.

The engagement effectiveness metric addresses a critical gap in current cybersecurity practices by providing a quantifiable means to assess the success of cognitive warfare operations. This standardized measurement enables cybersecurity professionals, social media platforms, and governments to better monitor and mitigate the spread of malicious content. It can also inform real-time decision-making processes, such as flagging disinformation content that surpasses critical engagement thresholds, prompting further investigation or removal of content.

While our research makes significant strides in understanding the engagement dynamics of cyber cognitive attacks, several opportunities remain for future exploration. Future work could enhance the metric by accounting for factors such as user diversity, bot activity, and the temporal evolution of engagements. Integrating this model with existing cybersecurity frameworks, such as the DISARM framework, could further bolster real-time detection and response strategies against cognitive warfare. Additionally, exploring how this metric can be applied to emerging platforms like TikTok and Reddit, or in the context of political and military operations, could yield new insights into the evolving landscape of cognitive attacks.

In conclusion, the proposed engagement effectiveness metric provides a vital foundation for advancing the study of cognitive warfare. By offering a quantifiable, adaptable approach to measuring disinformation engagement, this research paves the way for improved global cybersecurity resilience and more effective defense strategies against the growing threat of cognitive attacks.

\nocite{*}
\bibliography{dac_references}

\pagebreak
\section{Appendix: Data Availability}

We will maximize our work's scientific and community value by making it as open as possible. All code, data, and other materials needed to reproduce this project will be released publicly under an open-source license, and we will share all artifacts needed to reproduce this work.
This includes metric automation example code(s), charts, graphs, and other data as detailed in this paper.
\newpage

\section*{Appendix: Metric Automation Example Code}

The following Python code demonstrates how to automate the calculation of engagement effectiveness ($E$) using the weighted interaction model described in this paper. It includes configurable weights for views, likes, comments, and shares and can be extended to support additional interaction types.

\begin{verbatim}
# Example engagement data (including views)
posts_data = [
    {"post_id": 1, "views": 5000, "likes": 200, "comments": 50, "shares": 10},
    {"post_id": 2, "views": 100000, "likes": 3000, "comments": 400, "shares": 200},
    {"post_id": 3, "views": 100, "likes": 10, "comments": 2, "shares": 1},
]

# Define the weights for each interaction type
WEIGHTS = {
    "views": 0.1,
    "likes": 0.3,
    "comments": 0.7,
    "shares": 1.0
}

def calculate_engagement_effectiveness(post):
    # Calculate weighted interaction score
    return sum(post.get(k, 0) * w for k, w in WEIGHTS.items())

def assign_grade(e):  # Grade based on engagement effectiveness (E)
    if e >= 10000:
        return "A+", True
    elif e >= 1000:
        return "A", True
    elif e >= 100:
        return "B", False
    elif e >= 10:
        return "C", False
    elif e >= 4:
        return "D", False
    elif e >= 3:
        return "E", False
    else:
        return "F", False

def process_posts(posts):
    flagged_posts = []
    for post in posts:
        e = calculate_engagement_effectiveness(post)
        grade, should_flag = assign_grade(e)
        print(f"Post {post['post_id']} - E of {e:.2f} graded {grade}")
        if should_flag:
            flagged_posts.append(post["post_id"])
            print(f"Flagging post {post['post_id']} for review.")
    return flagged_posts

# Run analysis
flagged_content = process_posts(posts_data)

# Output review
if flagged_content:
    print(f"These posts are flagged for review: {flagged_content}")
else:
    print("No posts are flagged for review.")
\end{verbatim}

\end{document}